\documentclass[english, letterpaper, 11pt]{article}

\usepackage{setspace}
\usepackage[T1]{fontenc}
\usepackage[utf8]{inputenc}
\usepackage{babel}
\usepackage{blindtext}
\usepackage{cite}
\usepackage[margin=1in]{geometry}
\usepackage[margin=1in]{caption}
\usepackage{graphicx}
\graphicspath{{./fig/}}
\usepackage[cmex10]{amsmath}
\usepackage{amssymb}
\usepackage{array}
\usepackage[tight,footnotesize]{subfigure}
\usepackage{stfloats}
\usepackage{algorithm,algorithmic}
\usepackage{enumerate}
\usepackage{amsthm}
\usepackage{color}
\usepackage{mathrsfs}
\usepackage{multicol}
\usepackage{mathtools}
\usepackage{tikz}
\usepackage{pgfplots}
\usepackage{umoline}
\usepackage{authblk}

\providecommand{\keywords}[1]{\bigskip\textbf{\textit{Index terms---}} #1}

\newcommand{\eqdef}{\vcentcolon =}

\newcommand{\diag}{\operatorname{diag}}

\newcommand{\bbC}{\mathbb{C}}

\newcommand{\calR}{\mathcal{R}}
\newcommand{\calN}{\mathcal{N}}

\newcommand{\calX}{\mathcal{X}}

\newcommand{\calV}{\mathcal{V}}
\newcommand{\rom}[1]{\uppercase\expandafter{\romannumeral #1\relax}}
\newcommand{\transpose}{{\mathrm{T}}}
\newcommand{\vect}{\operatorname{vec}}
\newcommand{\spanof}{\operatorname{span}}

\newtheorem{theorem}{Theorem}[section]
\newtheorem{lemma}[theorem]{Lemma}

\newtheorem{claim}[theorem]{Claim}

\begin{document}

\title{Optimal Sample Complexity\\for Blind Gain and Phase Calibration\thanks{This work was supported in part by the National Science Foundation (NSF) under Grants CCF 10-18789 and IIS 14-47879.}}
\author[1]{Yanjun Li}
\author[2]{Kiryung Lee}
\author[1]{Yoram Bresler}
\affil[1]{Coordinated Science Laboratory and Department of Electrical and Computer Engineering}
\affil[2]{Coordinated Science Laboratory and Department of Statistics}
\affil[1,2]{University of Illinois, Urbana-Champaign}
\date{}
\maketitle

\doublespacing

\abstract
Blind gain and phase calibration (BGPC) is a structured bilinear inverse problem, which arises in many applications, including inverse rendering in computational relighting (albedo estimation with unknown lighting), blind phase and gain calibration in sensor array processing, and multichannel blind deconvolution. The fundamental question of the uniqueness of the solutions to such problems has been addressed only recently. In a previous paper, we proposed studying the identifiability in bilinear inverse problems up to transformation groups. In particular, we studied several special cases of blind gain and phase calibration, including the cases of subspace and joint sparsity models on the signals, and gave sufficient and necessary conditions for identifiability up to certain transformation groups. However, there were gaps between the sample complexities in the sufficient conditions and the necessary conditions. In this paper, under a mild assumption that the signals and models are generic, we bridge the gaps by deriving tight sufficient conditions with optimal sample complexities.

\keywords{uniqueness, blind gain and phase calibration, sensor array processing, inverse rendering, SAR autofocus, multichannel blind deconvolution}


\section{Introduction}
Blind gain and phase calibration (BGPC) is a bilinear inverse problem (BIP) that arises in many applications. It is the joint recovery of an unknown gain and phase vector $\lambda$ and signal vectors $\phi_1,\phi_2,\cdots,\phi_N$ given the entrywise product $Y=\diag(\lambda)\Phi$, where $\Phi=[\phi_1,\phi_2,\cdots,\phi_N]$. In inverse rendering \cite{Nguyen2013}, when the surface profile (3D model) of the object is known, the joint recovery of the albedo\footnote{Albedo, also known as reflection coefficient, is the ratio of reflected radiation from a surface to incident radiation upon it.} and the lighting conditions is a BGPC problem. In sensor array processing \cite{Paulraj1985}, if the directions of arrival of source signals are properly discretized using a grid, and the sensors have unknown gains and phases, the joint recovery of the source signals and the gains and phases of the sensors is a BGPC problem. In multichannel blind deconvolution (MBD) with the circular convolution model, the joint recovery of the signal and multiple channels is a BGPC problem.

In a previous paper \cite{Li2015}, we derived general necessary and sufficient conditions for identifiability in a bilinear inverse problem up to a transformation group, and applied these to BGPC to give identifiability results under several scenarios. The results were given in terms of sample complexities: the number of samples required for a unique solution. In particular, we considered the subspace constraint and joint sparsity constraint scenarios for the signals, and derived sufficient conditions for the identifiability up to scaling (or other groups of equivalence transformations). We also gave necessary conditions in the form of tight lower bounds on sample complexities. We showed that the sufficient conditions and the necessary conditions coincide in some cases, and analyzed the gaps in other cases. We also presented conjectures on how to bridge the gaps.

In this paper, we prove one of the posed conjectures. In the subspace constraint scenario, we assume that the subspace model and the signals are generic. Then we show that the sample complexity in the necessary condition is actually sufficient for almost all signals. Therefore, the sample complexity is optimal. We also generalize this result to the joint-sparsity case, and derive a sample complexity that is almost optimal.

The rest of the paper is organized as follows. We introduce the problem setup and summarize our previous results in the rest of this section. In Section \ref{sec:main}, we state and prove the main results: the (almost) optimal sample complexities for BGPC with subspace or with joint sparsity constraints. We conclude in Section \ref{sec:conclusions} with some discussion.

\subsection{Notations}
Before proceeding to the problem statement, we state the notations that will be used throughout the paper. We use upper-case letters $A$, $X$ and $Y$ to denote matrices, and lower-case letters to denote vectors. The diagonal matrix with the elements of vector $\lambda$ on the diagonal is denoted by $\diag(\lambda)$. The vector formed by a concatenation of the columns of $X$ is denoted by $\vect(X)$. We use $I_n$ and $F_n$ to denote the identity matrix and the discrete Fourier transform (DFT) matrix of size $n\times n$. Unless otherwise stated, all vectors are column vectors. The dimensions of all vectors and matrices are made clear in the context. The Kronecker product of two matrices is denoted by $\otimes$. The entrywise product is denoted by $\odot$. The range space of the conjugate transpose of a matrix $D$ is denoted by $\calR^*(D) = \calR(D^*)$, and the nullspace of $D$ is denoted by $\calN(D)$. The orthogonal complement of a subspace $\calV$ is denoted by $\calV^\perp$. Given a vector $x\in\bbC^n$, $\spanof(x)$ denotes the one dimensional subspace of $\bbC^n$ spanned by $x$, and $x^\perp$ denotes its orthogonal complement.

We use $j,k$ to denote indices, and $J,K$ to denote index sets. If a matrix or a vector has dimension $n$, then an index set $J$ is a subset of $\{1,2,\cdots,n\}$. We use $|J|$ to denote the cardinality of $J$, and $J^c$ to denote its complement. We use superscript letters to denote subvectors or submatrices. Thus, $x^{(J)}$ represents the subvector of $x$ consisting of the entries indexed by $J$, with the scalar $x^{(j)}$ representing the $j$th entry of $x$. The submatrix $A^{(J,K)}$ has size $|J|\times |K|$ and consists of the entries indexed by $J\times K$. Borrowing the colon notation from MATLAB, the vector $A^{(:,k)}$ represents the $k$th column of matrix $A$.

\subsection{Problem Statement}

\emph{Blind gain and phase calibration (BGPC)} is the following constrained bilinear inverse problem (BIP) given the measurement $Y=\diag(\lambda_0)\Phi_0$:
\begin{align*}
\text{Find}~~&(\lambda,\Phi),\\
\text{s.t.}~~&\diag(\lambda)\Phi = Y,\\
& \lambda \in \Omega_\Lambda,~\Phi \in \Omega_\Phi,
\end{align*}
where $\lambda \in \Omega_\Lambda\subset\bbC^{n}$ is the unknown gain and phase vector, and $\Phi\in\Omega_\Phi\subset\bbC^{n\times N}$ is the signal matrix. In this paper, we impose no constraints on $\lambda$, i.e., $\Omega_\Lambda = \bbC^n$.
As for the matrix $\Phi$, we impose subspace or joint sparsity constraints. In both scenarios, $\Phi$ can be represented in the factorized form $\Phi=AX$, where the columns of $A\in \bbC^{n\times m}$ form a basis or a frame (an overcomplete dictionary), and $X\in \Omega_\calX\subset \bbC^{m\times N}$ is the matrix of coordinates. The constraint set becomes $\Omega_\Phi=\{\Phi=AX:X\in\Omega_\calX\}$. Under some mild conditions\footnote{Under a subspace constraint, $A$ is required to have full column rank. Under a joint sparsity constraint, $A$ is required to satisfy the spark condition \cite{Donoho2003}. Both conditions are satisfied by a generic $A$.} on $A$, the uniqueness of $\Phi$ is equivalent to the uniqueness of $X$. For simplicity, we treat the following problem as the BGPC problem from now on.
\begin{align*}
\text{(BGPC)}\qquad\text{Find}~~&(\lambda,X),\\
\text{s.t.}~~&\diag(\lambda)AX = Y,\\
& \lambda \in \bbC^n,~X \in \Omega_\calX.
\end{align*}

Next, we elaborate on the scenarios considered in this paper:

(\rom{1}) \emph{Subspace constraints.} The signals represented by the columns of $\Phi$ reside in a low-dimensional subspace spanned by the columns of $A$. The matrix $A$ is tall ($n>m$) and has full column rank. The constraint set is $\Omega_\calX=\bbC^{m\times N}$.
	
In inverse rendering \cite{Nguyen2013}, the columns of $Y=\diag(\lambda)\Phi$ represent images under different lighting conditions, where $\lambda$ represents the unknown albedos,\footnote{In inverse rendering, albedos are real and positive. We ignore this extra information here for simplicity.} and the columns of $\Phi$ represent the intensity maps of incident light. The columns of $A$ are the first several spherical harmonics extracted from the 3D model of the object. They form a basis of the low-dimensional subspace in which the intensity maps reside.
	
Multichannel blind deconvolution (MBD) with the circular convolution model also falls into this category. The measurement $Y^{(:,j)}=\diag(\lambda)\Phi^{(:,j)}$ can be also written as:	
\[
F_n^{-1}Y^{(:,j)}=(F_n^{-1}\lambda)\circledast(F_n^{-1}\Phi^{(:,j)}),
\]
where $\circledast$ denotes circular convolution, and $F_n^{-1}$ is the inverse DFT. The vector $\lambda$ represents the DFT of the signal, the columns of $\Phi$ represent the DFT's of the impulse responses of the channels, and the columns of $Y$ represent the DFT's of the channel outputs. The columns of $F_n^{-1}A$ form a basis for the low-dimensional subspace in which the impulse responses of the channels reside. For example, when the multiple channels are FIR filters that share the same support $J$, they reside in a low-dimensional subspace whose basis is $F_n^{-1}A=I^{(:,J)}$. By symmetry, the roles of signals and channels can be switched.
In channel encoding, when multiple signals are encoded by the same tall matrix $E$, they reside in a low-dimensional subspace whose basis is $F_n^{-1}A=E$. In this case, the vector $\lambda$ represents the DFT of the channel.
	
(\rom{2}) \emph{Joint sparsity constraints.} The columns of $\Phi$ are jointly sparse over a dictionary $A$, where $A$ is a square matrix ($n=m$) or a fat matrix ($n<m$). The constraint set $\Omega_\calX$ is \[\Omega_\calX=\{X\in\bbC^{m\times N}: \text{$X$ has at most $s$ nonzero rows}\}.\]
In other words, the columns of $X$ are jointly $s$-sparse.
	
In sensor array processing with uncalibrated sensors, the vector $\lambda$ represents unknown gains and phases of the sensors, and the columns of $\Phi$ represent snapshots captured at different time instants, assuming unit gain and zero phase for all sensors. Consider a scene with radiating sources whose positions (directions of arrival in the far-field scenario) are discretized, using a grid of $m$ positions. Then each column of $A\in\bbC^{n\times m}$ represents the array response to a single source at one position on the grid. With only $s<m$ unknown sources, each column of $\Phi$ is the superposition of the same $s$ columns of $A$. Hence the columns of the source matrix $X$ are jointly $s$-sparse. 

If the impulse responses in MBD are jointly sparse over the dictionary $F_n^{-1}A$, then as argued in the subspace constraints case, the vector $\lambda$, the columns of $\Phi$, and the columns of $Y$ represent the DFT's of the signal, the impulse responses, and the channel outputs, respectively. By symmetry, the roles of signals and channels can be switched. For example, in hyperspectral imaging, image samples at different frequencies in the light spectrum are likely to share the same discontinuities, and be jointly sparse over the same dictionary. If all image samples are corrupted with the same blurring kernel, then the deblurring procedure is a BGPC problem with joint sparsity constraints.

Synthetic aperture radar (SAR) autofocus \cite{Morrison2009} is a special multichannel blind deconvolution problem, where $X$ represents the SAR image and $A=F$ is the 1D DFT matrix. The entries in $\lambda$ represent the phase error in the Fourier imaging data, which varies only along the cross-range dimension.\footnote{In SAR autofocus, the entries of the phase error $\lambda$ have unit moduli. We ignore this extra information here for simplicity.} If we extend the coverage of the image by oversampling the Fourier domain in the cross-range dimension, the rows of the image $X$ corresponding to the region that is not illuminated by the antenna beam will be zeros. Thus, the SAR image $X$ can be modeled as a matrix with jointly sparse columns.

In the rest of this paper, we address the identifiability in the above BGPC problem. For BGPC, the constraint sets $\Omega_\Lambda$ and $\Omega_\calX$ are cones -- they are closed under scalar multiplication. For any nonzero scalar $\sigma$, the pairs $(\lambda_0,X_0)$ and $(\sigma\lambda_0,\frac{1}{\sigma}X_0)$ map to the same $Y$ and hence are non-distinguishable. We say that this problem suffers from scaling ambiguity. The set $\{(\sigma\lambda_0,\frac{1}{\sigma}X_0): \sigma\neq 0\}$ is an equivalence class of solutions generated by a group of scaling transformations. We say that the solution $(\lambda_0,X_0)$ is identifiable up to scaling if every solution to BGPC is a scaled version of $(\lambda_0,X_0)$ in that equivalent class. In this paper, we answer the following question: under what conditions is the solution $(\lambda_0,X_0)$ unique up to scaling?

Our results are stated in terms of sample complexities, which are the numbers of data samples or measurements needed for unique recovery of the solutions. They are given by inequalities describing the conditions that need to be satisfied by the problem parameters, $n$, $m$, $s$, and $N$. The numbers $n$ and $m$ denote the length of the signals and the dimension of the subspace in which they are assume to reside, in the subspace constraint scenario. The sparsity level $s$ is the number (out of $m$) of nonzero rows of $X$ in the joint sparsity scenario. Finally, the number of signals captured (number of columns of $Y$ and $\Phi$) is denoted by $N$. Table \ref{tab1} summarizes what these parameters represent in the applications.
Since it is often difficult to acquire a large number of signals, it is desirable to have sample complexities that requires small $N$.

\begin{table}[htbp]
\newcolumntype{L}[1]{>{\raggedright\let\newline\\\arraybackslash\hspace{0pt}}m{#1}}
\newcolumntype{C}[1]{>{\centering\let\newline\\\arraybackslash\hspace{0pt}}m{#1}}
\newcolumntype{R}[1]{>{\raggedleft\let\newline\\\arraybackslash\hspace{0pt}}m{#1}}
\centering
\begin{tabular}{L{0.5cm} L{4.0cm} L{4.0cm} L{6.0cm} L{0cm}}
\hline\hline
& \multicolumn{1}{c}{Inverse Rendering} & \multicolumn{1}{c}{Sensor Array Processing} & \multicolumn{1}{c}{MBD} &\\[5pt]\hline
$n$ & \# pixels & \# sensors & Length of the signal &\\[5pt]\hline
$m$ & \# spherical harmonics & \# positions on the grid & Dimension of the channel subspace \newline\emph{(subspace constraint)} &\\[15pt]\hline
$s$ & \Midline{$\qquad$$\qquad$} & \# sources & Channel sparsity level \newline \emph{(joint sparsity constraint)} &\\[15pt]\hline
$N$ & \# images & \# snapshots & \# channels &\\[5pt]
\hline\hline
\end{tabular}
\caption{Summary of problem parameters}
\label{tab1}
\end{table}

\subsection{Related Work}

The structure of the BGPC problem arises in many signal processing applications. In each of these, the problem formulation and treatment were tailored to the application. For example, Nguyen et al. \cite{Nguyen2013} showed a sufficient condition for unique inverse rendering. Morrison et al. \cite{Morrison2009} proposed an algorithm for SAR autofocus and showed a necessary condition for their algorithm. Both problems fall into the category of BGPC problems with subspace constraints.

In a previous paper \cite{Li2015}, we addressed the identifiability of all BGPC problems in a common framework. We first considered BGPC with a subspace constraint, and no additional structure for the matrix $A$. For BGPC with a joint sparsity constraint, we considered the recovery of sparse signals, for which the matrix $A$ is the DFT matrix, and piecewise constant signals, for which the matrix $A$ is the product of the DFT matrix and a matrix whose columns form a basis for piecewise constant signals. In all these cases, we derived both sufficient conditions and necessary conditions for identifiability. 

A limitation of the previous work \cite{Li2015}, is that the sample complexities in the sufficient conditions are suboptimal. For example, for BGPC with a subspace constraint, the sample complexity in the sufficient condition is $N\geq m$. However, the necessary condition says that the sample complexity only needs to satisify  $N\geq \frac{n-1}{n-m}$. This less demanding sample complexity coincides with the bound obtained by counting the number of degrees of freedom and the number of measurements, and also agrees with the empirical phase transition \cite{Li2015}. The sufficient condition for identifiability in BGPC with a joint sparsity constraint at sparsity level $s$ suffers from similar suboptimality: the sufficient condition is $N\geq s$, versus the necessary condition $N\geq \frac{n-1}{n-s}$. In this paper, we show that the less demanding sample complexities are actually sufficient for almost all matrices $A$ and $X$.


\section{Main Results}\label{sec:main}


\subsection{BGPC with a Subspace Constraint} \label{sec:subspace}
We first consider identifiability in BGPC with a subspace constraint. The measurement in the following problem is $Y=\diag(\lambda_0)AX_0$. The known matrix $A\in\bbC^{n\times m}$ is tall ($n>m$). Hence the columns of $\Phi = AX$ reside in a low-dimensional subspace. The corresponding constraint sets are $\Omega_\Lambda = \bbC^n$ and $\Omega_\calX=\bbC^{m\times N}$, hence the problem is unconstrained with respect to $\lambda$ and $X$, and takes the form:
\begin{align*}
\text{Find}~~&(\lambda,X),\\
\text{s.t.}~~&\diag(\lambda)AX = Y,\\
& \lambda \in \bbC^n,~X \in \bbC^{m\times N}.
\end{align*}

In previous work \cite{Li2015}, we showed that $N\geq m$ is sufficient to guarantee identifiability when $A$, $\lambda_0$, and $X_0$ are generic. However, numerical experiments show that when $\frac{n-1}{n-m}\leq N\leq m$, the solution can still be identifiable (See \cite[Section 3.3]{Li2015}). In this section, we explore the regime where $\lambda_0$, $X_0$, and $A$ are generic, and $\frac{n-1}{n-m}\leq N\leq m$. We prove the following sufficient condition for the identifiability of $(\lambda_0,X_0)$ up to scaling.
\begin{theorem}\label{thm:subspace}
In the BGPC problem with a subspace constraint, if $n>m$ and $\frac{n-1}{n-m}\leq N\leq m$, then for almost all $\lambda_0\in\bbC^n$, almost all $X_0\in\bbC^{m\times N}$, and almost all $A\in\bbC^{n\times m}$, the pair $(\lambda_0,X_0)$ is identifiable up to an unknown scaling.
\end{theorem}

The sample complexity required by this theorem, $N\geq \frac{n-1}{n-m}$, is much less demanding than the condition $N\geq m$ in our previous results (\cite[Theorem 3.3 and Corollary 3.4]{Li2015}). In fact, this sample complexity is optimal, since it matches the sample complexity in the necessary condition (see \cite[Proposition 3.5]{Li2015}). It suggests that if $m\leq \frac{n}{2}$, i.e., the dimension of the subspace is less than half the ambient dimension, then $N=2$ signals are sufficient to recover $(\lambda_0,X_0)$ uniquely. This result provides a favorable bound for real world applications. For example, the typical dimension of the intensity map subspace in inverse rendering is $m=9$, which is really small when compared to the size of the images (e.g., $n=256\times 256=2^{16}$). Therefore, two images under different lighting conditions is all that is needed for the solution to be unique.
We will prove this result in Section \ref{sec:proofsubspace}.

When the sample complexity is achieved, for almost all $\lambda_0$, $X_0$, and $A$, the solution $(\lambda_0,X_0)$ is unique up to scaling. In other words, this result is violated only for $(\lambda_0,X_0,A)$ on a subset of $\bbC^n\times \bbC^{m\times N} \times \bbC^{n\times m}$ that has Lebesgue measure zero. If $(\lambda_0,X_0,A)$ is a random variable, following a distribution that is absolutely continuous with respect to the Lebesgue measure, then the solution to BGPC is identifiable up to scaling with probability $1$. 

As shown later in the proof of Theorem \ref{thm:subspace}, the identifiability hinges on the following conditions:
\begin{enumerate}
	\item There are no zero rows in $AX_0$, and all the entries of $\lambda_0$ are nonzero.
	\item The matrix in \eqref{eq:null}, which is a function of $A$ and $X_0$, has full column rank. 
\end{enumerate}
For a given combination of $\lambda_0$, $X_0$, and $A$, we can test whether the above conditions are satisfied, to determine whether the solution $(\lambda_0,X_0)$ is unique up to scaling. Moreover, the degenerate set of $(\lambda_0,X_0,A)$ that fails the test, is an algebraic variety, which is not dense in the ambient space. In real-world applications, $\lambda_0$ and $AX_0$ represent natural signals. Unless nature is malicious, they will not belong to the particular lower-dimensional manifold of degeneracy.


\subsection{BGPC with a Joint Sparsity Constraint}
Next, we consider identifiability in BGPC with a joint sparsity constraint. The measurement is $Y=\diag(\lambda_0)AX_0$. The columns of $A\in\bbC^{n\times m}$ form a basis or frame for the signals. There are $s$ nonzero rows in $X_0$. The problem of recovering $(\lambda_0,X_0)$ subject to this constraint is stated as follows:
\begin{align*}
\text{Find}~~&(\lambda,X),\\
\text{s.t.}~~&\diag(\lambda)AX = Y,\\
& \lambda \in \bbC^n,~X \in \{X\in\bbC^{m\times N}: \text{ the columns of $X$ are jointly $s$-sparse}\}.
\end{align*}

In previous work \cite{Li2015}, sufficient conditions for the uniqueness of the solution to the above problem were derived for some special cases (e.g., $A=F$). A sample complexity $N\geq s$ was established as sufficient for these special cases. However, when $\lambda_0$, $X_0$, and $A$ are generic, a less demanding sufficient condition can be proved using essentially the same argument as in the proof of Theorem \ref{thm:subspace}. The proof is presented in Section \ref{sec:proofjointsparsity}.

\begin{theorem}\label{thm:jointsparsity}
In the BGPC problem with a joint sparsity constraint, if $n>2s$ and $\frac{n-1}{n-2s}\leq N\leq s$, then for almost all $\lambda_0\in\bbC^n$, almost all $X_0\in\bbC^{m\times N}$ with $s$ nonzero rows, and almost all $A\in\bbC^{n\times m}$, the pair $(\lambda_0,X_0)$ is identifiable up to an unknown scaling.
\end{theorem}

The sample complexity in this sufficient condition, $N\geq \frac{n-1}{n-2s}$ is far superior than the previous bound of $N\geq s$, when the sparsity level $s$ is much smaller than the ambient dimension $n$. For example, if $s< \frac{n}{4}$, then $N=2$ is sufficient. In sensor array processing, the number of sources $s$ is often much smaller than the number of sensors $n$. Therefore, we only need two snapshots to recover the unknown gains and phases uniquely. This is especially significant when the working conditions of the sensor array and/or the source locations vary over time. We can achieve higher temporal resolution by solving BGPC using fewer snapshots.


\section{Proof of the Main Results}

\subsection{Proof of Theorem \ref{thm:subspace}}\label{sec:proofsubspace}

First, BGPC is a bilinear inverse problem. Theorem 2.8 \cite{Li2015} stated equivalent conditions for identifiability in bilinear inverse problems up to some transformation groups. Specializing this result to the identifiability in BGPC up to scaling, we have the following lemma:
\begin{lemma}\label{lem:bip}
In BGPC, the pair $(\lambda_0, X_0)\in \Omega_\Lambda\times \Omega_\calX$ ($\lambda_0\neq 0,X_0\neq 0$) is identifiable up to scaling if and only if the following two conditions are met:
\begin{enumerate}
	\item If $\diag(\lambda_1)AX_1=\diag(\lambda_0)AX_0$ for some $(\lambda_1, X_1)\in \Omega_\Lambda\times \Omega_\calX$, then $X_1 = \sigma X_0$ for some nonzero $\sigma$.
	\item If $\diag(\lambda_1)AX_0=\diag(\lambda_0)AX_0$ for some $\lambda_1\in \Omega_\Lambda$, then $\lambda_1 = \lambda_0$.
\end{enumerate} 
\end{lemma}

We first show that Condition 2 holds: that is, if $X_0$ is given, then the recovery of $\lambda_0$ is unique. Note that for almost all matrices $A\in\bbC^{n\times m}$ and $X_0\in\bbC^{m\times N}$, there are no zero rows in the product $AX_0$. It follows that, if $\diag(\lambda_0)AX_0 = \diag(\lambda_1)AX_0$ for some $\lambda_1\in\bbC^n$, then $\lambda_1=\lambda_0$. 

By Lemma \ref{lem:bip}, to complete the proof, we only need to show that Condition 1 also holds for generic $\lambda_0$, $X_0$, and $A$.\footnote{We use arguments similar to those used for the proof of Theorem 4.2 in \cite{Li2015b}.} Suppose there exists $(\lambda_1,X_1)$ such that $\diag(\lambda_0)AX_0 = \diag(\lambda_1)AX_1$. Consider the $k$-th row on both sides of the equation, which can be written as
\[
(I_N \otimes A^{(k,:)})\vect(X_0) \lambda_0^k = (I_N \otimes A^{(k,:)})\vect(X_1) \lambda_1^k.
\]
Now, for almost all $\lambda_0$, $X_0$, and $A$, the left hand side is nonzero. Hence $\lambda_1$ and $X_1$ are nonzero. It follows that
\[
(I_N \otimes A^{(k,:)})\bigl(\vect(X_1)-\frac{\lambda_0^k}{\lambda_1^k}\vect(X_0)\bigr) = 0,
\]
and hence,
\[
\vect(X_1) \in \calN(I_N \otimes A^{(k,:)}) + \spanof(\vect(X_0)).
\]
Next, we project $\vect(X_1)$ onto the orthogonal complement of $\spanof(\vect(X_0))$. It follows that
\[
P_{\vect(X_0)^\perp} \vect(X_1) = \vect(X_1) - P_{\spanof(\vect(X_0))}\vect(X_1) \in  \calN(I_N \otimes A^{(k,:)}) + \spanof(\vect(X_0)).
\]
For linear vector spaces $\calV_1$ and $\calV_2$, $\calV_1+\calV_2 = (\calV_1^\perp \bigcap \calV_2^\perp)^\perp$. Using the fact that $\calN(I_N \otimes A^{(k,:)})^\perp = \calR^*(I_N \otimes A^{(k,:)})$, and $\spanof(\vect(X_0))^\perp=\vect(X_0)^\perp$, we have
\[
P_{\vect(X_0)^\perp}\vect(X_1)\in \bigl( \calR^*(I_N \otimes A^{(k,:)}) \bigcap \vect(X_0)^\perp \bigr)^\perp,\quad \text{for $k=1,2,\cdots,n$}.
\]
Taking note of the fact that $P_{\vect(X_0)^\perp}\vect(X_1)\in \vect(X_0)^\perp$, we have
\begin{equation}
P_{\vect(X_0)^\perp}\vect(X_1)\in \vect(X_0)^\perp \bigcap \biggl(\bigcap\limits_{k=1,2,\cdots,n}\bigl( \calR^*(I_N \otimes A^{(k,:)}) \bigcap \vect(X_0)^\perp \bigr)^\perp \biggr). \label{eq:intersect}
\end{equation}
Since
\begin{align*}
I_N \otimes A^{(k,:)} &=
\begin{bmatrix}
A^{(k,:)} & 0 & 0 & \cdots & 0 \\
0 & A^{(k,:)} & 0 & \cdots & 0 \\
0 & 0 & A^{(k,:)} & \cdots & 0 \\
\vdots & \vdots & \vdots & \ddots & \vdots \\
0 & 0 & 0 & \cdots & A^{(k,:)}
\end{bmatrix},\\
\vect(X_0)^* &= 
\begin{bmatrix}
X_0^{(:,1)*} & X_0^{(:,2)*} & X_0^{(:,3)*} & \cdots & X_0^{(:,N)*}
\end{bmatrix},
\end{align*}
it is easy to verify that the intersection of the row space of $I_N \otimes A^{(k,:)}$ and the orthocomplement of $\vect(X_0)$ is
\[
\calR^*(I_N \otimes A^{(k,:)}) \bigcap \vect(X_0)^\perp = \calR^*\left(D(A^{(k,:)},X_0)\right),
\]
where the matrix $D(A^{(k,:)},X_0)\in\bbC^{(N-1)\times mN}$ is a function of $A^{(k,:)}$ and $X_0$:
\[
D(A^{(k,:)},X_0) = 
\begin{bmatrix}
-A^{(k,:)}X_0^{(:,2)} & A^{(k,:)}X_0^{(:,1)} & 0 & \cdots & 0\\
-A^{(k,:)}X_0^{(:,3)} & 0 & A^{(k,:)}X_0^{(:,1)} & \cdots & 0\\
\vdots & \vdots & \vdots & \ddots & \vdots\\
-A^{(k,:)}X_0^{(:,N)} & 0 & 0 & \cdots & A^{(k,:)}X_0^{(:,1)}
\end{bmatrix}\otimes A^{(k,:)}.
\]
For generic matrices $A$ and $X_0$, $D(A^{(k,:)},X_0)$ has full row rank, which is $N-1$. By \eqref{eq:intersect},
\begin{equation}
P_{\vect(X_0)^\perp}\vect(X_1) \in \calN\left(
\begin{bmatrix}
\vect(X_0)^*\\
D(A^{(1,:)},X_0)\\
D(A^{(2,:)},X_0)\\
\vdots\\
D(A^{(n,:)},X_0)
\end{bmatrix}
\right).
\label{eq:null}
\end{equation}
We have the following claim, which we will prove in Section \ref{sec:proveclaim}.
\begin{claim}\label{cla:rank}
For almost all $X_0$ and $A$, if $n>m$ and $\frac{n-1}{n-m}\leq N\leq m$, then the matrix in \eqref{eq:null} has full column rank, which is $mN$.
\end{claim}
Given this claim, for almost all $X_0$ and $A$, $P_{\vect(X_0)^\perp}\vect(X_1)=0$. Therefore, $X_1$ resides in the $1$-dimensional subspace in $\bbC^{m\times N}$ spanned by $X_0$, i.e., $X_1 = \sigma X_0$. Recall that $X_1$ is nonzero, hence $\sigma\neq 0$, establishing Condition 2 in Lemma \ref{lem:bip}, thus proving Theorem \ref{thm:subspace}.

\subsection{Proof of Claim \ref{cla:rank}}\label{sec:proveclaim}
We prove that the matrix in \eqref{eq:null} has full column rank for almost all $X_0$ and $A$ that satisfy $n>m$ and $\frac{n-1}{n-m}\leq N\leq m$. By the definition of matrix $D(A^{(k,:)},X_0)$, we have $D(A^{(k,:)},X_0)\vect(X_0)=0$. Hence the first row $\vect(X_0)^*$ is orthogonal to the rest of the rows in the matrix in \eqref{eq:null}. Therefore, we only need to show the following matrix has rank $mN-1$ for almost all $X_0$ and $A$:
\[
D(A,X_0) = 
\begin{bmatrix}
D(A^{(1,:)},X_0)\\
D(A^{(2,:)},X_0)\\
\vdots\\
D(A^{(n,:)},X_0)
\end{bmatrix}\in\bbC^{n(N-1)\times mN}
\]
The rank of $D(A,X_0)$ is at most $mN-1$, since all its rows are orthogonal to $\vect(X_0)^*$. We only need to show the rank is at least $mN-1$ for almost all $A$ and $X_0$.

By a basic result in algebraic geometry, the rank of $D(A,X_0)$ is at least $mN-1$ for almost all $A$ and $X_0$, if the rank is $mN-1$ for at least one choice of $A$ and $X_0$. The rest of the proof is an explicit construction of $A$ and $X_0$ that satisfies this rank.

\begin{figure}[htpb]
\centering
\includegraphics[width=0.8\columnwidth]{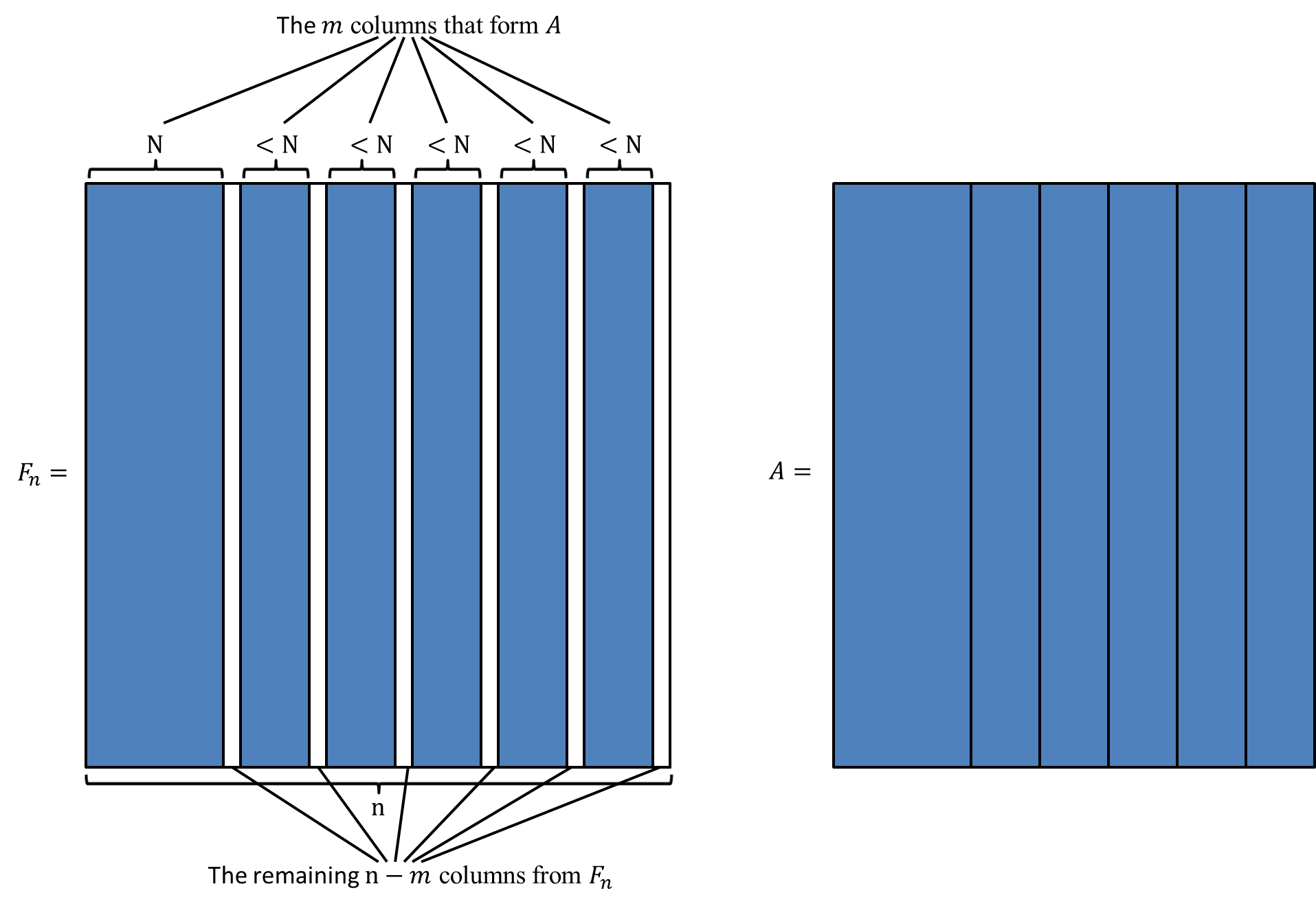}%
\caption{Picking $m$ columns from $F_n$ as the columns of $A$}%
\label{fig:pickcolumns}%
\end{figure}

The matrix $X_0$ is a tall matrix ($N\leq m$), hence we can choose $X_0$ as the first $N$ columns of $I_m$. The matrix $A$ is also tall ($n>m$), hence we can choose $A$ as a subset of $m$ columns from $F_n$. The first $N$ columns are $A^{(:,1:N)}=F_n^{(:,1:N)}$. We pick $m-N$ columns out of $F_n^{(:,N+1:n)}$ as $A^{(:,N+1:m)}$ in a manner such that there are no blocks of consecutive $N$ columns except for the first $N$ columns. Hence the columns $F_n^{(:,N+1)}$ and $F_n^{(:,n)}$ must not be picked.\footnote{Because of the circular nature of the DFT matrix, the first column and the last column of $F_n$ are also considered ``consecutive''.} This can be demonstrated by Figure \ref{fig:pickcolumns}. This can be done because $(n-m)N\geq n-1$.

Given this choice of $X_0$ and $A$,
\[
D(A^{(k,:)},X_0) =
\begin{bmatrix}
-\alpha^{k-1} & 1 & 0 & \cdots & 0\\
-\alpha^{2(k-1)} & 0 & 1 & \cdots & 0\\
\vdots & \vdots & \vdots & \ddots & \vdots\\
-\alpha^{(N-1)(k-1)} & 0 & 0 & \cdots & 1
\end{bmatrix} \otimes A^{(k,:)},
\]
where $\alpha = e^{-\frac{2\pi\sqrt{-1}}{n}}$. We can view $D(A,X_0)$ as a block matrix with $n$ blocks, one on top of the other. Each block itself is a block matrix with $(N-1)\times N$ blocks.

Consider the left null vector $w\in\bbC^{n(N-1)}$ of the matrix $D(A,X_0)$. Suppose
\[
w = [w_{1,1}, w_{1,2}, \cdots, w_{1,N-1}, w_{2,1}, w_{2,2},\cdots,w_{2,N-1},\cdots,w_{n,1},w_{n,2},\cdots,w_{n,N-1}]^\transpose,
\]
and $w^*D(A,X_0) = 0$. Then we have
\begin{align}
\sum_{k=1,2,\cdots,n}\left(\sum_{j=1,2,\cdots,N-1}\alpha^{j(k-1)}\overline{w_{k,j}}\right) A^{(k,:)} = 0, \label{eq:leftnull1}\\
\sum_{k=1,2,\cdots,n}\overline{w_{k,j}} A^{(k,:)} = 0, \text{ for $j= 1,2,\cdots,N-1$.} \label{eq:leftnull2}
\end{align}
In order to show that $D(A,X_0)$ has rank $mN-1$, we need to prove that there are exactly $M \eqdef n(N-1)-(mN-1)=nN-mN-n+1$ linearly independent left null vectors $w$. This number is greater than or equal to zero because $N\geq \frac{n-1}{n-m}$. Consider the following matrix:
\[
W = 
\begin{bmatrix}
w_{1,1} & w_{1,2} & \cdots & w_{1,N-1} \\
w_{2,1} & w_{2,2} & \cdots & w_{2,N-1} \\
\vdots & \vdots & \ddots & \vdots \\
w_{n,1} & w_{n,2} & \cdots & w_{n,N-1} \\
\end{bmatrix}.
\]
By \eqref{eq:leftnull2}, the columns of $W$ are orthogonal to the columns of $A$. Recall that the columns of $A$ are a subset of the columns of $F_n$. We use $A_\perp\in\bbC^{n\times(n-m)}$ to denote the matrix whose columns are the complement set of columns, i.e., the remaining $n-m$ columns in $F_n$ that are not picked. Then $W=A_\perp Q$ for some $Q\in\bbC^{(n-m)\times(N-1)}$. Next, we show that there are exactly $M$ linearly independent matrices $Q$ such that $W=A_\perp Q$ satisfies \eqref{eq:leftnull1}. 

Consider the following vector $v\in\bbC^{n}$ whose entries are the coefficients in \eqref{eq:leftnull1}:
\begin{align}
v &\eqdef
\begin{bmatrix}
\sum\limits_{j=1,2,\cdots,N-1}\alpha^{-j\cdot 0}w_{1,j}\\
\sum\limits_{j=1,2,\cdots,N-1}\alpha^{-j\cdot 1}w_{2,j}\\
\vdots\\
\sum\limits_{j=1,2,\cdots,N-1}\alpha^{-j(n-1)}w_{n,j}
\end{bmatrix} \nonumber\\
&= \sum_{j=1,2,\cdots,N-1} F_n^{(:,n+1-j)}\odot W^{(:,j)} \nonumber\\
&= \sum_{i=1,2,\cdots,n-m}~\sum_{j=1,2,\cdots,N-1} \bigl(F_n^{(:,n+1-j)}\odot A_\perp^{(:,i)}\bigr)Q^{(i,j)}. \label{eq:doublesum}
\end{align}
The entrywise product of two columns in $F_n$ is still a column in $F_n$. In particular, if $j_2>j_1$, then $F_n^{(:,n+1-j_1)}\odot F_n^{(:,j_2)} = F_n^{(:,j_2-j_1)}$. Therefore, for every $i$ and $j$, $F_n^{(:,n+1-j)}\odot A_\perp^{(:,i)}$ is a column in $F_n$. The vector $v$ is a linear combination of the columns in $F_n$.\footnote{There can be repeated columns in this sum.} By \eqref{eq:leftnull1}, $v$ is also orthogonal to the columns in $A$. Therefore, there exists a vector $p\in\bbC^{n-m}$ such that
\begin{equation}
v = A_\perp p = \sum_{i=1,2,\cdots,n-m} A_\perp^{(:,i)} p^{(i)}. \label{eq:singlesum}
\end{equation}
By \eqref{eq:doublesum} and \eqref{eq:singlesum}, we have
\begin{equation}
\sum_{i=1,2,\cdots,n-m} A_\perp^{(:,i)} p^{(i)} - \sum_{i=1,2,\cdots,n-m}~\sum_{j=1,2,\cdots,N-1} \bigl(F_n^{(:,n+1-j)}\odot A_\perp^{(:,i)}\bigr)Q^{(i,j)} = v - v = 0. \label{eq:tworep}
\end{equation}
Recall that $F_n^{(:,N+1)}$ and $F_n^{(:,n)}$ are not picked for $A$, and hence belong to $A_\perp$. Based on the way we partition $F_n$ into $A$ and $A_\perp$, at least one column in any $N$ consecutive columns in $F_n^{(:,N+1:n)}$ must belong to $A_\perp$. In other words, there are at most $N-1$ columns in $A$ whose original indices in $F_n$ are between the original indices of two adjacent columns in $A_\perp$. The only exception is that between $F_n^{(:,n)}$ and $F_n^{(:,N+1)}$, which are adjacent columns in $A_\perp$ because they are the last and first columns, there are $N$ columns $F_n^{(:,1:N)}$. Therefore, if we consider the $N-1$ columns $F_n^{(:,n+1-j)}\odot A_\perp^{(:,i)}$ ($j=1,2,\cdots,N-1$), they ``fill the gap'' and include all the columns in $A$ whose indices are between the indices of $A_\perp^{(:,i-1)}$ and $A_\perp^{(:,i)}$, with the only exception that $F_n^{(:,1)}$ is not included in any of these. Hence,
\[
\left\{A_\perp^{(:,i)}: 1\leq i\leq n-m\right\}\bigcup\left\{ F_n^{(:,n+1-j)}\odot A_\perp^{(:,i)}: 1\leq i\leq n-m, 1\leq j\leq N-1 \right\} = \left\{ F_n^{(:,j)}: 2\leq j\leq n \right\},
\]
i.e., all the columns in the sum of \eqref{eq:tworep} span a subspace of dimension $n-1$. Hence, there are $(n-m)+(n-m)(N-1)-(n-1) = nN-mN-n+1=M$ linearly independent choices of coefficients $[\vect(Q)^\transpose,p^\transpose]^\transpose$. We denote these linearly independent vectors by $[\vect(Q_k)^\transpose,p_k^\transpose]^\transpose$, $k=1,2,\cdots,M$. 

Next we prove that $Q_1,Q_2,\cdots,Q_M$ are linearly independent. We argue by contradiction. Suppose they are linearly dependent, and there exists $\beta_1,\beta_2,\cdots,\beta_M$ such that
\begin{equation}
\sum_{k=1,2,\cdots,M} \beta_k Q_k = 0. \label{eq:dependence}
\end{equation}
Then,
\begin{align*}
&  A_\perp \biggl(\sum_{k=1,2,\cdots,M} \beta_k  p_k \biggr) \\
= & \sum_{k=1,2,\cdots,M} \beta_k  A_\perp p_k\\
= & \sum_{k=1,2,\cdots,M}~\beta_k \sum_{i=1,2,\cdots,n-m}~\sum_{j=1,2,\cdots,N-1} \bigl(F_n^{(:,n+1-j)}\odot A_\perp^{(:,i)}\bigr) Q_k^{(i,j)}\\
= &  \sum_{i=1,2,\cdots,n-m}~\sum_{j=1,2,\cdots,N-1} \bigl(F_n^{(:,n+1-j)}\odot A_\perp^{(:,i)}\bigr) \biggl(\sum_{k=1,2,\cdots,M}\beta_k Q_k^{(i,j)}\biggr)\\
=0.
\end{align*}
The second equation follows from \eqref{eq:tworep}, and the last equation follows from \eqref{eq:dependence}. Since the matrix $A_\perp$ has full column rank, we have
\begin{equation}
\sum_{k=1,2,\cdots,M} \beta_k p_k = 0. \label{eq:dependence2}
\end{equation}
Equations \eqref{eq:dependence} and \eqref{eq:dependence2} suggest that $[\vect(Q_k)^\transpose,p_k^\transpose]^\transpose (k=1,2,\cdots,M)$ are linearly dependent, which causes a contradicition. Therefore, $Q_1,Q_2,\cdots,Q_M$ are linearly independent. There exist exactly $M$ linearly independent left null vectors for $D(A,X_0)$. Therefore, $D(A,X_0)$ has rank $mN-1$ for the special choice of $A$ and $X_0$, which completes the proof.

\subsection{Proof of Theorem \ref{thm:jointsparsity}}\label{sec:proofjointsparsity}
First, by the same argument as in the proof of Theorem \ref{thm:subspace}, if $X_0$ is given, the recovery of $\lambda_0$ is unique. Again by Lemma \ref{lem:bip}, we only need to show that for generic $\lambda_0$, $X_0$, and $A$, if there exists $(\lambda_1,X_1)$ such that $\diag(\lambda_0)AX_0=\diag(\lambda_1)AX_1$, then $X_1=\sigma X_0$ for some nonzero $\sigma$.

We start by fixing the supports of $X_0$ and $X_1$. Suppose $\diag(\lambda_0)AX_0=\diag(\lambda_1)AX_1$, and $J_0$ and $J_1$ are the row supports (the index set on which the rows of a matrix are nonzero) of $X_0$ and $X_1$, respectively, and $|J_0|=|J_1|=s$. Then focus on the following equation, containing the nonzero rows of $X_0$ and $X_1$:
\[
\diag(\lambda_0)A^{(:,J_0\bigcup J_1)}X_0^{(J_0\bigcup J_1,:)}=\diag(\lambda_1)A^{(:,J_0\bigcup J_1)}X_1^{(J_0\bigcup J_1,:)}.
\]
Obviously, the cardinality of the set $J_0\bigcup J_1$ is at most $2s$. Let $\ell=|J_0\bigcup J_1|\leq 2s$. We can show that $X_1^{(J_0\bigcup J_1,:)} = \sigma X_0^{(J_0\bigcup J_1,:)}$ for some nonzero $\sigma$, following the same steps as in the proof of Theorem \ref{thm:subspace}, with Claim \ref{cla:rank} replaced by the following claim:
\begin{claim}\label{cla:rank2}
For almost all $X_0$ with row support $J_0$ and almost all $A$, if $n>2s\geq \ell$ and $\frac{n-1}{n-2s}\leq N\leq s$, then the following matrix has full column rank, which is $\ell N$:
\begin{equation}
\begin{bmatrix}
\vect(X_0^{(J_0\bigcup J_1,:)})^*\\
D(A^{(1,J_0\bigcup J_1)},X_0^{(J_0\bigcup J_1,:)})\\
D(A^{(2,J_0\bigcup J_1)},X_0^{(J_0\bigcup J_1,:)})\\
\vdots\\
D(A^{(n,J_0\bigcup J_1)},X_0^{(J_0\bigcup J_1,:)})
\end{bmatrix}. \label{eq:js_eq1}
\end{equation}
\end{claim}
The proof of Claim \ref{cla:rank2} uses arguments similar to those in the proof of Claim \ref{cla:rank}: an explicit construction of $A^{(:,J_0\bigcup J_1)}$ and $X_0^{(J_0\bigcup J_1,:)}$ that satisfies a rank condition described below. Here, we cannot choose every entry of $X_0^{(J_0\bigcup J_1,:)}$ freely, since it has only $s$ nonzero rows. Let $Q$ be an $\ell\times\ell$ permutation matrix, such that the first $s$ rows of $QX_0^{(J_0\bigcup J_1,:)}$ are nonzero. Then we apply the construction of $A$ and $X$ in the the proof of Claim \ref{cla:rank}, to $A^{(:,J_0\bigcup J_1)}Q^{-1}$ and $QX_0^{(J_0\bigcup J_1,:)}$. For example, we choose $X_0^{(J_0\bigcup J_1,:)}$ such that $QX_0^{(J_0\bigcup J_1,:)}$ is the first $N\leq s$ columns of $I_\ell$. Then by the proof of Claim \ref{cla:rank}, the following matrix has full column rank $\ell N$:
\begin{equation}
\begin{bmatrix}
\vect(QX_0^{(J_0\bigcup J_1,:)})^*\\
D(A^{(1,J_0\bigcup J_1)}Q^{-1},QX_0^{(J_0\bigcup J_1,:)})\\
D(A^{(2,J_0\bigcup J_1)}Q^{-1},QX_0^{(J_0\bigcup J_1,:)})\\
\vdots\\
D(A^{(n,J_0\bigcup J_1)}Q^{-1},QX_0^{(J_0\bigcup J_1,:)})
\end{bmatrix}. \label{eq:js_eq2}
\end{equation}
We complete the proof of Claim \ref{cla:rank2} by making the following observation: \eqref{eq:js_eq2} is a permutation of the columns of \eqref{eq:js_eq1}, and the two matrices have the same rank.

We continue the proof of Theorem \ref{thm:jointsparsity}. We have established that $X_1^{(J_0\bigcup J_1,:)} = \sigma X_0^{(J_0\bigcup J_1,:)}$ for some nonzero $\sigma$. Recall that the other rows of $X_0$ and $X_1$ are zero. Hence $X_1 = \sigma X_0$. Therefore, for almost all $\lambda_0$ and $A$, and almost all $X_0$ whose row support is $J_0$, the solution $(\lambda_1,X_1)$, for which the support of $X_1$ is $J_1$, satisfies that $X_1=\sigma X_0$ and $\lambda_1=\frac{1}{\sigma}\lambda_0$. There are a finite number of choices for the supports $J_0$ and $J_1$, ${m\choose s}^2$ choices to be exact. Therefore, we can complete the proof by enumerating over all possible choices for $J_0$ and $J_1$. 

\section{Conclusions}\label{sec:conclusions}
In this paper, we addressed the identifiability of the BGPC problem with subspace or joint sparsity constraint, up to scaling. We gave sufficient conditions for identifiability that feature optimal (or almost optimal) sample complexities. These results are for generic vectors or matrices, and are violated only for a set of Lebesgue measure zero.

We did not address the stability of BGPC in this paper. The regime under which the problem can be solved stably is an interesting open problem.


\end{document}